\documentclass[12pt,a4paper]{article}
\usepackage[latin1]{inputenc}
\usepackage{amsmath}
\usepackage{pstricks}
\usepackage{amssymb}
\usepackage{bm}
\usepackage{pgf,pgfarrows,pgfnodes,pgfautomata,pgfheaps}
\usepackage{url}
\newcommand{\ben}{\begin{eqnarray}}
\newcommand{\bbn}{\begin{eqnarray}}
\newcommand{\een}{\end{eqnarray}}
\newcommand{\be}{\begin{eqnarray*}}
\newcommand{\bb}{\begin{eqnarray*}}
\newcommand{\ee}{\end{eqnarray*}}

\newcommand{\parno}{\par\noindent}
\begin{document}
%
%
\begin{center}
{\bf\Large Closed form solutions of\\[0.5em] measures of systemic risk } \\[1em]
\textit{Manfred Jäger-Ambro\.zewicz\footnote{Comments are appreciated: Manfred.Jaeger-Ambrozewicz@HTW-Berlin.de. Webpage \url{www.htw-berlin.de/organisation/?typo3state=persons&lsfid=8754} or \url{www.mathfred.de}. I would like to thank Brice Hakwa (University Wuppertal), Thomas Hartmann-Wendels (University Cologne), Barbara Rüdiger-Mastandrea (University Wuppertal), Eric Schaanning (ETH) and the participants of the Rhein Wupper Seminar on Financial Engineering and Risk Management and the Fachsysmposium ``Das neue Gesicht der Regulierung im Finanzmarkt'' of the University Cologne for comments.}\\
Hochschule für Technik und Wirtschaft Berlin\\
HTW University of Applied Science Berlin}\\
11-Nov-2012 \\[2em]
\end{center}
\begin{abstract}
This paper derives -- considering a Gaussian setting -- closed form solutions of the statistics that Adrian and Brunnermeier \cite{adrian2009} and Acharya et al. \cite{acharya2009} have suggested as measures of systemic risk to be attached to individual banks. The statistics equal the product of statistic specific $\beta$-coefficients with the mean corrected Value at Risk. Hence, the measures of systemic risks are closely related to well known concepts of financial economics. Another benefit of the analysis is that it is revealed how the concepts are related to each other. Also, it may be relatively ``easy'' to convince the regulators to consider a closed form solution, especially so if the statistics involved are well known and can easily be communicated to the financial community.  
\end{abstract}
\setlength{\baselineskip}{1.35em}
\section{Introduction}
Value at Risk (VaR) and Expected Shortfall (ES) are established measures of riskiness of e.g. financial institutions (e.g. McNeil et al. \cite{embrechts}). However, these numbers measure the riskiness of an institution in ``isolation''. The financial crisis taught everybody that it is not a good idea to measure the riskiness of financial institutions as if these institutions could be analyzed in isolation. Indeed, systemic risk, how to measure it and how to allocate responsibilities is currently an energetically researched topic of economics (e.g. Acharya et al. \cite{acharya2009}, \cite{acharya2010}, Adrian and Brunnermeier \cite{adrian2009}, \cite{adrian2010}, Dreh\-mann and Tarashev \cite{drehmann2011}, Galati and Moessner \cite{galati}, Gau\-thier, Lehar and Souissi \cite{gauthier}, Huang, Zhou and Zhu \cite{huang}, Tarashev, Borio, Tsatsaronis \cite{tarashev2011}, Zhou \cite{zhou2010}). Hellwig \cite{hellwig} is a early contribution stressing the importance of systemic risk analysis. Crockett \cite{crockett} is frequently cited as the seminal contribution stressing the need to augment a microprudential analysis with a macroprudential one. Concerning the approach of this paper Lehar \cite{lehar} should be considered as seminal. He stresses the similarity of the macroprudential approach to the quantitative analysis of a risk manager supervising a portfolio of risks.   
\par This paper is based on two influential recent contributions: Adrian and Brunnermeier \cite{adrian2009}, \cite{adrian2010} and Acharya et al. \cite{acharya2009}, \cite{acharya2010}. Adrian and Brunnermeier have stressed the need to analyze systemic risk and have done so in a very practical manner. They explicitly develop a statistic that could play a practical role when fixing regulatory capital. Adrian and Brunnermeier argue very convincingly that incentives of banks would be distorted if regulation were based on Value at Risk not taking into account the systemic risk linked to a bank. Indeed, a bank might be incentivated to assume to much systemic risk if this risk is not addressed by regulation. Acharya et al. \cite{acharya2009} elaborate on the necessity to use regulation to internalize externalities related to systemic risk. They argue that two kinds of externalities are relevant (Acharya et al. \cite{acharya2009}, p. 286). First, a failing bank may trigger spillover effects that the bank does not take into account when chosing its strategy and business model. Indeed, in case of a crisis ``liquidity spiral, leading to depressed asset prices and a hostile funding environment, pulling others down and thus leading to further price drops, funding illiquidity, and so on'' (see also Brunnermeier and Pedersen \cite{brunnermeier2009}. Second, in case of a systemic crisis, there is no private solver -- an institution that could for example buy a failing bank -- as all private actors are suffering in a systemic crisis. Hence, the cost of solving the problem are socialized (e.g. via a subsidized mergers or subsidized capital injections).   
\par Adrian and Brunnermeier \cite{adrian2010} consider the Value of Risk of a group $A$ of financial institutions given that a specific institution $i$ has hit its $\text{VaR}^i$ ($\text{CoVaR}^{Ai}$). To measure the systemic risk to be attached to bank $i$, they suggest
\bb \Delta \text{CoVaR}^{Ai} &=& \text{CoVaR}^{Ai} - \text{CoVaRm}^{Ai} 
\\ &=& \text{VaR}(X_A | X_i = \text{VaR}(X_i) )  - \text{VaR}(X_A | X_i = \text{median}(X_i) ). \ee 
Obviously, $\text{CoVaR}$ depends on the dependencies that the other institutions have with the state of the institution $i$, especially the dependencies at the left (risky) tail. 
\par In this paper the approach of Adrian and Brunnermeier is studied in a simple stochastic Gaussian setting. Within this framework the intuition of $\Delta \text{CoVaR}$ can easily be grasped and a closed form solution can be derived. The approach is very transparent as it can easily be linked to standard concepts of risk management viz. $\beta$-coefficient and Value at Risk (VaR). 
The approach is as follows: We consider a system of financial institutions $S = \left\{i\right\} \cup A$, where we single out one of these institutions (viz. the institution $i$). We derive the very simple closed form representation of the systemic risk attached to bank $i$ (Delta Collateral Value at Risk):
\bb \Delta \text{CollVaR}^{Ai} &=& - \Phi^{-1} (\alpha) \frac{\Sigma_{Ai}}{\sqrt{\Sigma_{i}}} = \beta_{Ai} \, \text{VaR}^{\text{mean}}(X_i), \\
\Delta \text{CollVaR}^{Ai} &=& \text{CoVaR}^{Ai} - \text{VaR}^A(X_A|X_i = \mathbf{E}(X_i)), \\
\beta_{Ai} &=& \frac{\text{cov} (X_i , X_A) }{ \text{var} (X_i)}, \\ 
\text{VaR}^{\text{mean}} (X_i) &=& \text{VaR}(X_i) - \mathbf{E} (X_i),   \ee
where $\alpha$ is the VaR-threshold, $\Phi$ is the standard normal distribution function and $\Sigma_{Ai}$ is the covariance of the returns of $i$ and $A$. We derive a series of closed form solutions of statistics related to systemic risk. Table 1 provides a summary. Note that there is only a minor difference to Adrian and Brunnermeier \cite{adrian2010}. We use the expected value instead of the median to define the condition of the conditional Value at risk.         
\par We also have a closed form solution if instead of Value of Risk Expected Shortfall is considered: 
\bb \Delta \text{CollES}^{Ai} &=& - \frac{\phi (\Phi^{-1} (\alpha))}{1-\alpha} \frac{\Sigma_{Ai}}{\sqrt{\Sigma_{i}}} = \beta_{Ai} \, \text{ES}^{\text{mean}}(X_i) , \ee
where $\phi$ is the density of the standard normal distribution.  
\par Adrian and Brunnermeier \cite{adrian2010} analyze $\Delta \text{CoVaR}$ empirically using quantile regression, where explanatory variables are taken into account to produce forecasts of CoVaR, which is of crucial importance for a regulator. The purpose of this paper is threefold: Improve intuition, offer closed form solutions in a specific stochastic framework and reveal how the concepts are related to each other.  
\par The structure of the paper is as follows. In section 2 the Gaussian setting is introduced and the closed form solutions for several statistics relevant for systemic risk are derived. It is shown that all statistics are closely related to well-known concepts of finance viz. $\beta$ and VaR. Also, the relationships between the statistics is derived. Section 3 summarizes.  
\section{Systemic Risk in a Gaussian setting}
\subsection{The stochastic framework and the statistics analyzed}
We focus on the following problem: There is a system of banks, we single out one of these banks and we aim to measure the systemic risk to be attached to the bank singled out. Such a perspective is quite natural for a supervisor, who actually will study the risk of bank on its own  first and augment this narrow risk analysis by a systemic risk analysis. The aim of the paper is to derive simple formulae of quantitative measures of systemic risk to be attached to the bank focused. We denote the bank focused by $i$. $A$ refers to the complete banking system without the bank $i$, i.e. $S = \left\{i\right\} \cup A$ is the system under investigation.     
\par In the following a vector $(X_i , X_A)'$ of a bank-$i$ respectively group-$A$ related statistic will be considered. Deliberately, we leave indefinite what statistic is considered; one may think of Profit/Loss, Return, Assets, Equity, etc. 
\\ \parno \textbf{Maintained Assumption:} \textit{We assume that $X_i , X_A$ are jointly Gaussian with expected values $\mu_i , \mu_A$ and var\-iance-covariance matrix}  
\bb \Sigma = 
\left( \begin{array}{cc}   \Sigma_i &  \Sigma_{Ai}   \\
 \Sigma_{Ai} 	 & \Sigma_A  \end{array} \right).
\ee
In the following and without further mentioning, we are going to assume that this assumption holds. Of course it is this assumption that allows us to derive closed form solutions. It is well known that many statistics in finance are not Gaussian (for a survey see Rockinger et al. \cite{rockinger2006}). Hence, the closed form solutions derived later should be used with much care. But the benefits of transparency and intuition that result from a closed form solution justify to also study the Gaussian case. 
\par The following lemma offers a closed form solution of the conditional expected value and of the conditional variance (see McNeil et al. \cite{embrechts}, p. 68).     
\\ \parno \textbf{Lemma 1:} \textit{The conditional expected value of $X_A$ given that $X_i = x$ is 
\bb {\bf E} (X_A |X_i = x) =  \mu_A + \frac{\Sigma_{Ai}}{\Sigma_{i}} (x-\mu_i) \ee
and the conditional variance equals }
\bb {\bf VAR} (X_A |X_i = x) = \Sigma_{A} - \frac{\Sigma_{Ai}^2}{\Sigma_{i}}. \ee
Note, that the expected value of $X_A$ depends linearly on realization $x$ of $X_i$, whereas the conditional variance does not depend on $x$. Note, that the variance of the conditioned random variable is smaller than its unconditional variance if $X_i$ an $X_A$ are positively correlated (let us call this effect ``variance reduction by conditioning'').   
\par The second lemma of great use for our purpose endows us with a closed form solution of the Value at Risk of a normally distributed variable (see McNeil et al. \cite{embrechts}, p. 39)). 
\\ \parno \textbf{Lemma 2:} \textit{The Value at Risk of a normally distributed variable $X$ is given by}  
\bb \textrm{VaR}(X) = \mu - \varphi  \cdot \sqrt{\Sigma}  \, ,\,\, \varphi = \Phi^{-1} (\alpha), \ee
\textit{where $\Phi^{-1}$ is the inverse of the standard normal distribution and $\alpha$ is the probability threshold used to define the Value at Risk.}
\\ \parno \textbf{Remark:} For clarification, the parameter $\alpha$ is relatively close to 1 (say 0.999) -- and hence $\varphi$ is a positive real number ($\varphi = 3.09$ if $\alpha = 0.999$). 
\\ \parno \textbf{List of Statistics:} The aim of this paper is to study statistics that measure systemic risk to be attached to a bank. We will derive closed form solutions of three such statistics and also study the relationships between these statistics: 
\bb \Delta \text{CollVaR}^{Ai} = \text{VaR} (X_A | X_i = \text{VaR} (X_i) )  - \text{VaR} (X_A | X_i = \mathbf{E} (X_i) ), \\
 \Delta \text{CondVaR}^{Si} = \text{VaR} (X_S | X_i = \text{VaR} (X_i) ) - \text{VaR} (X_S | X_i = \mathbf{E} (X_i) ),\\
 \Delta \text{ContrVaR}^{iS} = \text{VaR} (X_i | X_S = \text{VaR} (X_S) ) - \text{VaR} (X_i | X_S = \mathbf{E} (X_S) ). 
 \ee
In each case a stressed situation is compared with an unstressed situation. The stressed situation is modeled by using the VaR in the condition of the conditional Value of Risk, whereas the unstressed situation is modeled by using the expected value to fix the condition of the conditional VaR. Note, that by studying the difference of conditioned VaRs the variance reduction effect mentioned below lemma 1 cancels. This is indeed very important, as otherwise misleading statistics would be generated. 
\par At the margin note, that the difference to $\Delta$CoVaR of Adrian and Brunnermeier is merely the definition of the unstressed situation. Whereas in Adrian and Brunnermeier \cite{adrian2010} the median is used to define the condition, here the expected value is used. In a Gaussian setting of course the difference disappears as median and expected value are the same.         
\subsection{The closed form solution}
\par We first apply the two lemmata to derive a closed form solution of the VaR of $A$ given that bank $i$ has hit its VaR (we have to be careful here: $\Sigma_i$ denotes the variance and hence $\sqrt{\Sigma_i}$ is the standard deviation needed in the formula for VaR.), i.e. of $\text{CoVaR}^{Ai} = \text{VaR}(X_A|X_i=\text{VaR}(X_i))$.
We obtain 
\bb \text{CoVaR}^{Ai} &=& \mu_A + \frac{\Sigma_{Ai}}{\Sigma_{i}} \left( \mu_i - \sqrt{\Sigma_i} \cdot \varphi - \mu_i \right) - \varphi \sqrt{ \Sigma_{A} - \frac{\Sigma_{Ai}^2}{\Sigma_{i}} } \\ 
&=& \mu_A - \varphi \frac{\Sigma_{Ai}}{ \sqrt{\Sigma_{i}} } 
- \varphi \sqrt{ \Sigma_{A} - \frac{\Sigma_{Ai}^2}{\Sigma_{i}} }, 
\ee
The sum of the first two terms is the conditional expected value and the third is the conditional standard deviation. Lemma 1 is used to calculate the conditional VaR of $X_A$. Lemma 2 is used twice. First when calculating conditional VaR of $X_A$ and second when calculating the VaR of $X_i$. We are now in the position to calculate the closed form solution of    
\bb \Delta \text{CollVaR}^{Ai} &=& \text{CoVaR}^{Ai} - \text{CoVaRe}^{Ai},  \ee
where
\bb \text{CoVaRe}^{Ai} &=& \text{VaR} (X_A | X_i = \mathbf{E} (X_i) ) \\
                  &=& \mu_A + \frac{\Sigma_{Ai}}{ \Sigma_{i}  } ( \mu_i  - \mu_i) 
- \varphi \sqrt{   \Sigma_{A} - \frac{\Sigma_{Ai}^2}{ \Sigma_{i} }  } \\
&=& \mu_A - \varphi \sqrt{   \Sigma_{A} - \frac{\Sigma_{Ai}^2}{ \Sigma_{i} }  }.
\ee
We observe that most terms -- viz. the conditional variance and the unconditional expected value -- cancel out if we calculate the difference $\text{CoVaR}^{Ai}-\text{CoVaRe}^{Ai}$, so that we obtain 
\bb \Delta \text{CollVaR}^{Ai} =   - \varphi \frac{ \Sigma_{Ai} }{\sqrt{\Sigma_i}} 
= - \Phi^{-1} (\alpha) \frac{ \Sigma_{Ai} }{\sqrt{\Sigma_i}} \ee
and mutatis mutandis 
\bb \Delta \text{CollES}^{Ai} = -  \frac{\phi (\Phi^{-1} (\alpha))}{1-\alpha} \frac{ \Sigma_{Ai} }{\sqrt{\Sigma_i}}. \ee
\\ \parno
\textbf{Proposition 1:} \emph{Delta Collateral Value of Risk equals}  
\bb \Delta \text{CollVaR}^{Ai} &=& \left( \frac{\Sigma_{Ai}}{\Sigma_i} \right) \cdot (- \varphi  \sqrt{\Sigma_i}) \\ &=&
\beta_{Ai} \cdot \text{VaR}^{\text{mean}}(X_i). \ee
\emph{The interpretation is as follows: Delta Collateral Value at Risk equals the regression coefficient} $\beta_{Ai} = \left( \frac{\Sigma_{Ai}}{\Sigma_i} \right) $ \emph{times the impulse} $\text{VaR}^{\text{mean}}(X_i)$.\textit{ In other words: The calamities of bank i measured by its own mean-corrected Value at Risk $\text{VaR}^{\text{mean}}(X_i)$ are translated into systemic calamities via the regression coefficient $\beta_{Ai}$.}   
\\ \parno \textbf{Remark 1:} If $i$ is in a stressed situation $A$ is also pulled down (if $i$ and $A$ are correlated). The statistic $\Delta \text{CollVaR}^{Ai}$ captures this effect actually through the change in expected value:  
\bb \Delta \text{CollVaR}^{Ai} &=& \mathbf{E} ( X_A | X_i = \text{VaR}(X_i) ) - \mu_A. \ee
Also note that 
\bb \Delta \text{CollVaR}^{Ai} &=&  \beta_{Ai} \cdot \text{VaR}^{\text{mean}}(X_i) 
=  \frac{\text{cov}(X_i,X_A)}{ \text{var}(X_i)  } \cdot \text{VaR}^{\text{mean}}(X_i)
\\ &=&  - \frac{ \rho \cdot  \text{std}(X_i) \cdot  \text{std}(X_A)  }{ \text{std}(X_i)^2  } \cdot \varphi \cdot \text{std}(X_i)
= - \varphi \cdot \rho  \cdot \text{std}(X_A),    \ee
where $\rho = \text{corr}(X_i, X_A)$. Note, that the size of $i$ has no effect on $\Delta \text{CollVaR}^{Ai}$. $\Delta \text{CollVaR}^{Ai}$ is affected by the size of $A$ and the correlation. Hence, a small $i$ may be systemically relevant if $A$ is highly correlated with $i$. 
\par Note that in the process of supervising a bank, the statistic $\text{VaR}^{i,\text{mean}}$ is calculated anyway. All that is additionally needed to implement the formulae derived in this paper is the covariance $\text{cov} (X_i , X_S)$ (or $\beta_{Si}$ or $\beta_{iS}$). These statistics are relatively well understood and relatively easy to communicate to banks and to the market.   
\par So far we have considered the system $S = \{ i \} \cup A$. We have singled out the bank $i$ and studied the CoVaR attached to this bank. In this context CoVaR is best translated as Collateral Value at Risk: We consider the state of $A$ -- the others -- if $i$ is under stress. We thereby focus on the spillover (or the externality) attached to bank $i$ exerted on the group $A$. Obviously, it is also of interest to study the shape of the complete system $S$ if $i$ is under stress. In other words, we are as much interested to study the stochastic vector $(X_i , X_i + X_A)'$ as we are interested to analyze the vector $(X_i , X_A)'$. In the appendix we calculate variance-covariance matrix of $(X_i , X_i + X_A)' = (X_i , X_S)' $: 
\bb \Sigma^{iS} = 
\left( \begin{array}{cc}   \Sigma_i &  \Sigma_{iA} + \Sigma_i   \\
 \Sigma_{iA} + \Sigma_i 	 & \Sigma_i + 2\Sigma_{iA} + \Sigma_A   \end{array} \right).
\ee
Using this we find that the conditional Value at Risk of the system given that $X_i$ has hit the value $x$ is 
\bb \text{VaR} (X_S | X_i = x) = \mu_S +  \frac{ \Sigma_{iA} + \Sigma_i   }{ \Sigma_i } ( x - \mu_i) \\
 - \varphi \sqrt{  \Sigma_i + 2\Sigma_{iA} + \Sigma_A  - \frac{ (\Sigma_{iA} + \Sigma_i)^2  }{ \Sigma_i  } }
\ee
and consequently we have the 
\\ \parno \textbf{Proposition 2:} \emph{Delta Conditional Value at Risk equals}   
\bb \Delta \text{CondVaR}^{Si} &=& \text{VaR} (X_S|X_i = \text{VaR}(X_i))   - \text{VaR} (X_S|X_i = \mathbf{E}(X_i)) \\ 
 &=& - \varphi  \frac{ \Sigma_{iA} + \Sigma_i }{ \sqrt{\Sigma_i}  } \\ &=& \beta_{Ai} \text{VaR}^{\text{mean}}(X_i) + \text{VaR}^{\text{mean}}(X_i) = \beta_{Si} \text{VaR}^{\text{mean}}(X_i).  \ee
\emph{Again, this is intuitive. It means that the overall risk attached to bank $i$ consists of the bank's own risk plus its collateral risk. If bank $i$ hits its VaR the system is directly pulled down as $i$ is a member of the group: consequently we have the term $-\varphi \sqrt{\Sigma_i} = \text{VaR}^{\text{mean}}(X_i)$ (which is the mean corrected VaR). In addition to this we have an indirect negative effect (at least in the more likely case that $i$ and $A$ are positively correlated) as the other banks are simultaneously drawn down: $-\varphi \frac{\Sigma_{Ai}}{\sqrt{\Sigma_i}} =  \beta_{Ai} \text{VaR}^{\text{mean}}(X_i) $.} 
\\ \par So far we followed Adrian and Brunnermeier and considered the \textit{VaR of the system given that a bank has hit its VaR}. As an alternative we may consider 
\textit{the VaR of bank $i$ given that the financial system has hit its VaR}, i.e.  
\bb \text{VaR}( X_i | X_S = \text{VaR}(X_S)). \ee
This perspective is very close to the perspective that Acharya et al. \cite{acharya2009} and Brownless and Engle \cite{brownless2010} have recommended. They suggest to use a statistic that features prominently in risk management viz. the VaR-contribution: 
\bb \text{VaR-Contribution}_{iS} =  \mathbf{E} (X_i | X_S = \text{VaR}(X_S)) \ee
(Mc Neil, et. al. \cite{embrechts}, p. 258). VaR contribution defines the allocation principle if Value at Risk is used to allocate economic capital. The problem of capital allocation is to allocate the aggregate economic capital -- defined via a risk measure $\rho(L)$ -- to the risk factors $L_i, i=1,..,k$ of $L=\sum_{i=1..k} L_i$ (McNeil et al. \cite{embrechts}, p.  256ff). The idea of the VaR-Contribution: If the financial system has hit its VaR what is the contribution of $i$ to this loss. In other words, if we were to recapitalize a system (a portfolio) because of a systemic crisis (because of a joint loss of many assets) what is the share attributable to a specific institution (asset).     
\\ \parno \textbf{Proposition 3:}\textit{ The difference between the stressed and the normal situation is}
\bb
\Delta \text{ContrVaR}^{iS} &=&  \text{VaR}( X_i | X_S = \text{VaR}(X_S)) \\&& - \text{VaR}( X_i | X_S = \text{E}(X_S)) 
\\ &=& -\varphi  \frac{\Sigma_{iA} + \Sigma_i}{ \sqrt{\Sigma_i + 2\Sigma_{iA} + \Sigma_A} } =  -\varphi  \frac{\Sigma_{iA} + \Sigma_i}{ \sqrt{\Sigma_S} } \\
&=&  \frac{\Sigma_{iA} + \Sigma_i}{\Sigma_S} (-\varphi) \sqrt{\Sigma_S} \\&=& \beta_{iS} \text{VaR}^{\text{mean}}(X_S) .
\ee
where we use 
\bb \mathbf{E}(X_i | X_S = x) = \mu_i + \frac{ \Sigma_{iA} + \Sigma_i   }{ \Sigma_i + 2\Sigma_{iA} + \Sigma_A } (x - \mu_S), \\  
\text{\text{VaR}}(X_i | X_S = x) = \Sigma_i - \frac{ (\Sigma_{iA} + \Sigma_i)^2  }{ \Sigma_i + 2\Sigma_{iA} + \Sigma_A   }.
\ee
Hence
\bb
\text{VaR}( X_i | X_S = \text{VaR}(X_S)) &=&  \mu_i   - \frac{ \Sigma_{iA} + \Sigma_i   }{ \Sigma_i + 2\Sigma_{iA} + \Sigma_A }  \, \varphi\,   \sqrt{\Sigma_i + 2\Sigma_{iA} + \Sigma_A}   \\ 
&& - \varphi  \sqrt{\Sigma_i - \frac{ (\Sigma_{iA} + \Sigma_i)^2  }{ \Sigma_i + 2\Sigma_{iA} + \Sigma_A   }}  \\
&=& \mu_i  -\varphi  \frac{\Sigma_{iA} + \Sigma_i}{ \sqrt{\Sigma_i + 2\Sigma_{iA} + \Sigma_A} } 
\\&& - \varphi  \sqrt{\Sigma_i - \frac{ (\Sigma_{iA} + \Sigma_i)^2  }{ \Sigma_i + 2\Sigma_{iA} + \Sigma_A   }} 
\ee
Several remarks are appropriate: 
\\ \parno \textbf{Remark 2 (Aggregation):} A useful feature of $\Delta \text{ContrVaR}^{iS}$ is that the sum of the systemic risks attached to $i$ and $A$ equals aggregate risk. First observe
	\bb \beta_{iS} + \beta_{AS} =   \frac{\Sigma_{iA} + \Sigma_i}{\Sigma_S} + \frac{\Sigma_{iA} + \Sigma_A}{\Sigma_S}\\ = 
	 \frac{\Sigma_i + 2 \Sigma_{iA} + \Sigma_A}{\Sigma_S} = \frac{\Sigma_S}{\Sigma_S} = 1. \ee
Hence
	\bb  \Delta \text{ContrVaR}^{iS} + \Delta \text{ContrVaR}^{AS} = \text{VaR}^{\text{mean}}(X_S). \ee
\parno \textbf{Remark 3:} The VaR-Contribution -- a conditional expected value --  equals the unconditional expected value plus $\Delta \text{ContrVaR}$: 
\bb \text{VaR-Contribution}_{iS} = \mu_i + \Delta \text{ContrVaR}^{iS}. \ee
Indeed, using Lemma 1 we have   
\bb
\mathbf{E} (X_i | X_S = \text{VaR}(X_S)) &=&
 \mu_i -\frac{ \Sigma_{iA} + \Sigma_i   }{ \Sigma_i + 2\Sigma_{iA} + \Sigma_A } \, \varphi\,   \sqrt{\Sigma_i + 2\Sigma_{iA} + \Sigma_A} 
 \\ &=& \mu_i + \Delta \text{ContrVaR}^{iS} \ee
or 
\bb \Delta \text{ContrVaR}^{iS} = \mathbf{E} (X_i | X_S = \text{VaR}(X_S)) - \mu_i. \ee
Now, we can conpare this statistic with 
\bb \Delta \text{CollVaR}^{Ai} &=& \mathbf{E} ( X_A | X_i = \text{VaR}(X_i) ) - \mu_A \ee
In both cases a change of an expected value is considered. Whereas in the top-down approach one considers the change of expected value of $i$ given stress in the system, in the bottom up approach the perspective is reversed as one considers the expected value of the system given stress at $i$.     
\\ \parno \textbf{Remark 4:} It comes as no surprise that in a Gaussian framework risk related statistics refer to covariance and variance. Indeed, suppose that the standard deviation is used to measure the aggregate risk of $L=\sum_{i=1..k} L_i$. McNeil et al. \cite{embrechts}, p. 258 show that      
\bb AC_i^{\text{std}} = \frac{\text{cov}(L,L_i)}{\text{std}(L)} \ee
is the allocation principle for this risk measure. We observe  
\bb (- \varphi) \cdot AC_i^{\text{std}} = \Delta \text{ContrVaR}^{iS}. \ee
\parno \textbf{Remark 5:} Furthermore, note that \bb \sum_{i} \text{VaR-Contribution}_{iS} = \text{VaR}(X_S), \ee whereas \bb \Delta \text{ContrVaR}^{iS} + \Delta \text{ContrVaR}^{AS} = \text{VaR}^{\text{mean}}(X_S).\ee
This is as intuition suggests, as the statistics $\text{ContrVaR}^{xy}$ equal conditional expectations corrected with unconditional expectations. 
\\ \par The next result relates the statistics based on Adrian and Brunnermeier \cite{adrian2010} to that of Proposition 3. Indeed, the formula for $\Delta \text{ContrVaR}^{iS}$ closely resembles that of $\Delta \text{CondVaR}^{Si}$. The only difference is the denominator. Whereas it is $\frac{1}{\text{std} (X_i)}$ in case of $\Delta \text{CondVaR}^{Si}$ it is $\frac{1}{\text{std} (X_S)}$ in the  case of $\Delta \text{ContrVaR}^{iS}$. 
\\ \parno \textbf{Proposition 4:} \textit{The statistics $\Delta \text{CondVaR}^{Si}$ and  $\Delta \text{ContrVaR}^{iS}$ are closely related:} 
\bb \Delta \text{CondVaR}^{Si} = \left( \frac{\sqrt{\Sigma_S}}{\sqrt{\Sigma_i}} \right) \, \cdot \Delta \text{ContrVaR}^{iS}  \ee
\textit{or} 
\bb \frac{\Delta \text{CondVaR}^{Si}}{\sqrt{\Sigma_S}} =  \frac{\Delta \text{ContrVaR}^{iS} }{ \sqrt{\Sigma_i} } . \ee
Using Proposition 4 we can quickly find an additivity result for the Conditional Value at Risk.   
\\ \parno \textbf{Proposition 5:} \textit{The weighted sum of Conditional Values at Risk equals aggregated risk:}
\bb && \left( \frac{\sqrt{\Sigma_i}}{\sqrt{\Sigma_S}} \right) \Delta \text{CondVaR}^{Si} + \left( \frac{\sqrt{\Sigma_A}}{\sqrt{\Sigma_S}} \right) \Delta \text{CondVaR}^{SA} \\ &=&   \Delta \text{ContrVaR}^{iS} +  \Delta \text{ContrVaR}^{AS} = \text{VaR}^{\text{mean}}(X_S). \ee
This result mirrors the observation of remark 2. The sum of systemic risks attributed to the several banks equals mean corrected aggregate risk. However in case of the Conditional Value at Risk it is a \textit{weighted} sum, where the weights reflect the relative risk of the institutions. Proposition 5 shows how the top-down of Acharya et al. \cite{acharya2009} is related to the bottom-up approach of Adrian and Brunnermeier (see Drehmann and Tarashev \cite{drehmann2011} for a discussion of top-down and bottom-up).  
\section{Conclusion}
We have derived a battery of closed form solutions of statistics of systemic risk calculated in a Gaussian setting. The formulae allow us to relate the systemic risk statisics to well known concepts of financial economics viz. VaR and $\beta$-coefficient. We also derive a collection of result that reveal how the different statistics of systemic risk are related to each other. For sake of transparency a table of the formulae is provided. 
\\ \parno \textbf{Table 1: Collecting Results} 
\bb \Delta \text{CollVaR}^{Ai} &=& \text{VaR} (X_A | X_i = \text{VaR} (X_i) )   - \text{VaR} (X_A | X_i = \mathbf{E} (X_i) )\\ 
&=& \beta_{Ai} \text{VaR}^{\text{mean}}(X_i) = \rho \cdot \varphi \cdot \text{std}(X_A) \\
&=& \mathbf{E} ( X_A | X_i = \text{VaR}(X_i) ) - \mu_A  \\
 \Delta \text{CondVaR}^{Si} &=& \text{VaR} (X_S | X_i = \text{VaR} (X_i) ) - \text{VaR} (X_S | X_i = \mathbf{E} (X_i) )\\
&=& \beta_{Ai} \text{VaR}^{\text{mean}}(X_i) +  \text{VaR}^{\text{mean}}(X_i) = \beta_{Si} \text{VaR}^{\text{mean}}(X_i) \\
 \Delta \text{ContrVaR}^{iS} &=& \text{VaR} (X_i | X_S = \text{VaR} (X_S) ) - \text{VaR} (X_i | X_S = \mathbf{E} (X_S))\\
&=& \beta_{iS} \text{VaR}^{\text{mean}}(X_i) = \mathbf{E} (X_i | X_S = \text{VaR}(X_S)) - \mu_i \\
 \Delta \text{CondVaR}^{Si} &=& \left( \frac{\sqrt{\Sigma_S}}{\sqrt{\Sigma_i}} \right) \, \cdot \Delta \text{ContrVaR}^{iS} \\
\text{VaR}^{\text{mean}} &=& \Delta \text{ContrVaR}^{iS} +  \Delta \text{ContrVaR}^{AS} \\
\text{VaR}^{\text{mean}} &=& \left( \frac{\sqrt{\Sigma_i}}{\sqrt{\Sigma_S}} \right) \Delta \text{CondVaR}^{Si}  + \left( \frac{\sqrt{\Sigma_A}}{\sqrt{\Sigma_S}} \right) \Delta \text{CondVaR}^{SA}
 \ee
\section*{Appendix 1}
In addition to the stochastic vector $(X_i , X_A)'$ we here analyze the vector $(X_i , X_i + X_A)'$. Note that  
\bb X_i + X_A = ( 1 \quad  1 ) \left( \begin{array}{c}   X_i  \\ X_{A} \end{array} \right). \ee
Hence, we can use the formula (3.13) of McNeil et al. (\cite{embrechts}, 67) and obtain $\mathbf{E} (X_i + X_A)  = \mu_i + \mu_A$ and  
\bb \Sigma_S = ( 1 \quad  1 ) \Sigma \left( \begin{array}{c}   1  \\ 1 \end{array} \right) =
( 1 \quad  1 ) \left( \begin{array}{c}   \Sigma_i +  \Sigma_{Ai}   \\
 \Sigma_{Ai} 	 + \Sigma_A  \end{array} \right) \\ = \Sigma_i + 2\Sigma_{iA} + \Sigma_A.   \ee
Furthermore, the covariance of $X_S = X_A + X_i$ and $X_i$ can be calculated: 
\bb \text{cov}( X_A + X_i ,  X_i )   = \Sigma_{iA} + \Sigma_i.  \ee
Consequently the variance-covariance matrix of the vector $(X_i , X_i + X_A)'$ is given by  
\bb \Sigma = 
\left( \begin{array}{cc}   \Sigma_i &  \Sigma_{iA} + \Sigma_i   \\
 \Sigma_{iA} + \Sigma_i 	 & \Sigma_i + 2\Sigma_{iA} + \Sigma_A   \end{array} \right).
\ee
Equipped with these equations we can calculate the Value at Risk of the system $X_S$ 
\bb \text{VaR} (S) = \mu_S  - \varphi  \sqrt{\Sigma_i + 2\Sigma_{iA} + \Sigma_A} . \ee
We also obtain the conditional VaR using the formulae of McNeil et. al (\cite{embrechts}, 68). We note that 
\bb \mathbf{E} (X_S | X_i = x) = \mu_S  + \frac{ \Sigma_{iA} + \Sigma_i   }{ \Sigma_i } (x - \mu_i) \ee 
and 
\bb \text{var} (X_S | X_i = x) = \Sigma_i + 2\Sigma_{iA} + \Sigma_A  - \frac{ (\Sigma_{iA} + \Sigma_i)^2  }{ \Sigma_i  }. \ee
Consequently, we have that the conditional value at risk of $X_S$ if $X_i=x$ is given by
\bb \text{VaR} (X_S | X_i = x) = \mu_S +    \frac{ \Sigma_{iA} + \Sigma_i   }{ \Sigma_i } ( x - \mu_i ) \\
 - \varphi \sqrt{\Sigma_i + 2\Sigma_{iA} + \Sigma_A  - \frac{ (\Sigma_{iA} + \Sigma_i)^2  }{ \Sigma_i  }} 
\ee
and we have 
\bb  \text{CoVaR}^{iS} = \mu_S  - \frac{ \Sigma_{iA} + \Sigma_i   }{ \Sigma_i } \varphi \sqrt{\Sigma_i} 
 - \varphi \sqrt{\Sigma_i + 2\Sigma_{iA} + \Sigma_A  - \frac{ (\Sigma_{iA} + \Sigma_i)^2  }{ \Sigma_i  }} \\
= \mu_S - \varphi  \frac{\Sigma_{iA} + \Sigma_i}{\sqrt{\Sigma_i}}  
- \varphi \sqrt{\Sigma_i + 2\Sigma_{iA} + \Sigma_A  - \frac{ (\Sigma_{iA} + \Sigma_i)^2  }{ \Sigma_i  }}.
\ee

\begin{thebibliography}{}
\bibitem{acharya2009}
Acharya, V., Pedersen, L., Philippon, Z., Richardson, M.: Regulating Systemic Risk, In: Acharya, V. , Richardson, M. (eds): Restoring Financial Stability, pp. 283-303. Wiley, Hoboken (2009)
\bibitem{acharya2010}
Acharya, V., Pedersen, L., Philippon, T., Richardson, M.: Measuring Systemic Risk. Preprint (2010). \verb|http://pages.stern.nyu.edu/~sternfin/vacharya/public_html/Me|\\\verb|asuringSystemicRisk_final.pdf| 
\bibitem{adrian2009}
Adrian, T., Brunnermeier, M.: CoVaR. FRB of New York Staff Report No. 348 (2008). \verb|http://www.newyorkfed.org/research/staff_reports/sr348.pdf|
\bibitem{adrian2010}
Adrian, T., Brunnermeier, M.: CoVaR. Preprint (2010). \verb|http://www.princeton.edu/~markus/research/papers/CoVaR.pdf| 
\bibitem{brunnermeier2009}
Brunnermeier, M., Pedersen, L.: Market Liquidity and Funding Liquidity. Review of Financial Studies \textbf{22}(6), 2201-2238, (2009)
\bibitem{brownless2010}
Brownless, C., Engle, R.: Volatility, Correlation and Tails for Systemic Risk Measurement. Preprint (2011). \verb|http://ssrn.com/abstract=1611229| 
\bibitem{crockett}
Crockett, A.: Marrying the micro- and macroprudential dimensions of financial stability. BIS--Speeches (21. September 2000) (2000). \verb|http://www.bis.org/speeches/sp061005.htm|
\bibitem{drehmann2011}
Drehmann, M., Tarashev, N.: Systemic importance: some simple indicators. BIS Quarterly Review March 2011, 25--37 (2011)
\bibitem{lehar}
Lehar, A.: Measuring systemic risk: A risk management approach. Journal of Banking and Finance \textbf{29}(10), 2577-2603 (2009)
\bibitem{galati}
Galati, G., Moessner, R.: Macroprudential policy -- a literature review. BIS Working Papers No 337 (2011). \verb|http://www.bis.org/publ/work337.htm|
\bibitem{gauthier}
Gauthier, C., Lehar, A., Souissi, M.: Macroprudential Regulation and Systemic Capital Requirements. Bank of Canada Working Paper 2010-4 (2010) 
\bibitem{hellwig}
Hellwig, M.: Systemic Aspects of Risk Management in Banking and Finance. Schweizerische Zeit\-schrift für Volkswirtschaft und Statistik/Swiss Journal of Economics and Statistics, \textbf{131}, 723-737 (1995)
\bibitem{huang}
Huang, X., Zhou, H., Zhu, H.: Systemic Risk Contribution. Finance and Economics Discussion Series Division of Research and Statistics and Monetary Affairs Federal Reserve Board, Washington (D.C.) (2011)
\bibitem{engle}
Engle, R.: Anticipating Correlation. Princeton University Press, Princeton (2009).
\bibitem{embrechts}
McNeil, A., Frey, R., Embrechts, P.: Quantitative Risk Management. Princeton University Press, Princeton (2005).
\bibitem{rockinger2006}
Rockinger, M., Jondeau, E., Poon, S. H.: Financial Modeling under Non-Gaussian Distributions. Springer Verlag, Berlin, Heidelberg, New York (2006)
\bibitem{tarashev2011}
Tarashev, N., Borio, C., Tsatsaronis, K.: Attributing systemic risk to individual institutions. BIS Working Papers No 308 (2010)
\bibitem{zhou2010}
Zhou, C., Are banks too big to fail? Measuring systemic importance of financial institutions, International Journal of Central Banking 6(4), 205-250 (2010)
\end{thebibliography}
\end{document}